\documentclass[conference]{IEEEtran}

\usepackage{cite}
\usepackage{amsmath}
\usepackage{amssymb}
\usepackage{amsfonts}
\usepackage{amsthm}
\usepackage{graphicx}
\usepackage{textcomp}
\usepackage{enumitem}
\usepackage{subfigure}
\usepackage{footnote}

\usepackage{accents}
\usepackage{epsfig}
\usepackage{bm}
\usepackage[font={small}]{caption}
\usepackage{subcaption}
\usepackage{float}
\usepackage{algorithm}
\usepackage{algorithmicx}
\usepackage{algpseudocode}
\usepackage{xcolor}
\usepackage{comment}
\def\BibTeX{{\rm B\kern-.05em{\sc i\kern-.025em b}\kern-.08em
    T\kern-.1667em\lower.7ex\hbox{E}\kern-.125emX}}

\newtheorem{thm}{Theorem}

\newtheorem{lem}{Lemma}
\newtheorem{prop}{Proposition}

\newtheorem{defn}{Definition}
\newtheorem{exmp}{Example}

\newcommand{\C}{\mathbb{C}}
\newcommand{\Z}{\mathbb{Z}}
\newcommand{\F}{\mathcal{F}}

\newcommand{\V}{\mathcal{V}}

\newcommand{\be}{\begin{equation}}
\newcommand{\ee}{\end{equation}}
\newcommand{\beu}{\begin{equation*}}
\newcommand{\eeu}{\end{equation*}}
\newcommand{\ba}{\begin{aligned}}
\newcommand{\ea}{\end{aligned}}

\newcommand{\lba}{\left[ \begin{array}}
\newcommand{\ear}{\end{array} \right]}

\IEEEoverridecommandlockouts 

\begin{document}

\title{Factored Output Feedback Controller Synthesis with Locality Constraints for Spatially-Invariant Systems}

\author{\IEEEauthorblockN{Walden Marshall}
\IEEEauthorblockA{\textit{Department of Electrical, Computer \& Energy Engineering} \\
\textit{University of Colorado, Boulder}\\
walden.marshall@colorado.edu}
}

\maketitle

\begin{abstract}
    We consider $\mathcal{H}_2$ output feedback controller synthesis with pre-specified constraints on spatial communication distance (locality) for spatially-invariant systems using two factored controller frameworks: the system-level parameterization and the input-output parameterization. In our main result, we show that in both frameworks, output feedback controller synthesis with locality constraints can be formulated as a convex problem in finitely many transfer function variables, admitting the use of standard numerical solution techniques. The number of decision variables in the optimal controller design problem scales linearly with the distance of allowed communication. We also show that the optimal controller design problems for the system-level and input-ouptput parameterizations are equivalent for the chosen system of interest. We present numerical examples to illustrate the tradeoff between communication sparsity and performance.
\end{abstract}

\begin{IEEEkeywords}
    distributed control, infinite-dimensional systems, spatially-invariant systems, system-level synthesis
\end{IEEEkeywords}

\section{Introduction}
Optimal controller synthesis subject to structural communication constraints has long been of interest in the area of distributed control. In certain situations, e.g. funnel causality \cite{bamieh2005convex} and quadratic invariance \cite{rotkowitz2005characterization}, controller synthesis subject to structural constraints is a convex problem, but in general, it is nonconvex.\par
Some works \cite{rotkowitz2009closest,fazelnia2017relaxation,furieri2020sparsity} have focused on convex restrictions or relaxations  of the structurally constrained optimal controller design problem. The system-level (SL) parameterization \cite{doyle2017SLS,wang2019system,matni2017scalable,anderson2017realization}, the input-output (IO) parameterization \cite{furieri2019inputoutput,zheng2021equivalence,zheng2022numerical}, and network realization functions \cite{sabau2023network,sperila2023descriptor} are convex restrictions which factor a controller into distinct transfer function matrices and impose structural constraints on the factors. For SL and IO, these factors have a simple interpretation: they are closed-loop mappings from disturbances to state, measurement and control effort signals. By imposing structural constraints on these closed-loop mappings instead of on the controller transfer function matrix itself, SL and IO allow for synthesis of a controller with an implementation that respects network topology\cite{rantzer2019realizability}. Both structured controller transfer function matrices and controllers composed of structured factors can be implemented such that subcontrollers compute their action based only on information allowed by a pre-specified communication network, though design of structured controller transfer function matrices and design of structured factors are generally distinctly different problems \cite{jensen2020gap}.\par
In this paper, we use SL and IO to design structured closed-loop mappings for a plant comprised of a countably infinite number of identical subsystems which are coupled with other nearby subsystems. While real systems of interest have a finite number of subsystems, if the number of subsystems is large, approximating the whole system as having an infinite number of subsystems may be useful as it admits simpler computations through the use of the results in \cite{bamieh2002distributed}. For spatially-invariant systems, a structural constraint often considered \cite{jensen2018locality,jensen2020backstepping} is locality, where subcontrollers are resticted to utilize only signals from nearby spatial sites to compute their control action. \cite{jensen2018locality} shows that using SL and imposing locality constraints on the closed-loop mappings, optimal controller design for an infinite string of first-order subsystems with state feedback can be formulated as a convex problem in a finite number of transfer function decision variables. \cite{jensen2020backstepping} uses a backstepping approach to generalize these results to a subclass of higher order subsystems.\par
In this work, we offer another expansion to the work of \cite{jensen2018locality}. Our main results use \textit{both SL and IO} with locality-constrained closed-loop mappings to design \textit{output} feedback controllers for a class of \textit{second-order} subsystems. We show through an example that with both SL and IO, optimal controller design for spatially-invariant systems with locality constraints on the closed-loop factors can be formulated as a convex problem in finitely many transfer function variables, with the number of decision variables scaling linearly with the allowed spatial extent of closed-loop responses. We utilize a model-matching approach to find the true optimal closed-loop responses, rather than using the popular technique of restricting the closed-loop responses to have finite impulse response \cite{alonso2023MPC,anderson2017realization,zheng2022numerical}. In addition to providing true optimal performance, allowing infinite impulse response closed-loop mappings may admit lower order controller realizations compared to finite impulse response mappings.
\par
The paper is structured is follows. In Section II we give background on spatially-invariant systems, factored controllers and locality. In Section III we introduce the optimal control problem of interest. In Sections IV we state our main results: convex formulations of the optimal control problem with SL and IO. In Section V we provide proofs of our main results. In Section VI, we show the equivalence of the SL and IO optimal control problems. In Section VII, we present numerical results.

\section{Notation and Preliminaries}
We consider signals which are discrete in both time and space. Each signal is a function of two variables: either temporal index $t\in\Z^+$ or temporal frequency $z\in\C$, and either spatial index $k\in\Z$ or spatial frequency $\theta\in[0,2\pi)$.\par
We use over-accents and subscripts to convey temporal information about variables. Over-bar $(\bar{~})$ denotes that a variable is a function of $t$. Over-tilde $(\tilde{~})$ denotes that a variable is a function of $z$. Subscripts $(_t)$ and $(_z)$ indicate that the variable is evaluated at a single value of temporal index or frequency respectively. Absence of any of these notational devices indicates that whatever statement being made about the variables in question holds for any $t$ or any $z$.\par
Similarly, we use under-accents and subscripts to convey spatial information about variables. Under-bar $(\underbar{~})$ denotes that a variable is a function of $k$, while under-tilde $(\undertilde{~})$ denotes that a variable is a function of $\theta$. Subscripts $(_k)$ and $(_\theta)$ indicate that a variable is evaluated at a single spatial index or frequency respectively. Absence of any of these notational devices indicates that whatever statement being made about the variables in question holds for any $k$ or any $\theta$.\par
We denote multiple-input-multiple-output (MIMO) mappings between signals with upper case letters. We denote scalar mappings and vector-valued signals with lower case letters. We use superscripting to indicate the element of a signal or map, i.e. $\bar{\underline{x}}^m = (e^m)^T\bar{\underline{x}}$ and $\bar{\underline{g}}^{mn}=(e^m)^T\bar{\underline{G}}e^n$ where $e^j$ is the $j^{th}$ standard basis unit vector. \par
Signals which are functions of temporal or spatial frequency are computed using Fourier transforms as in \cite{jensen2018locality}. The Fourier transforms of $\bar{\underline{x}}$ are as follows:
\begin{itemize}
    \item Spatial Fourier transform:
    \be \label{eq:spacial_xform_def}
        \undertilde{\bar{x}} = \sum_{k\in\Z}\bar{x}_k e^{-ik\theta},
    \ee
    \item Temporal Fourier transform:
    \be \label{eq:temporal_xform_def}
        \underline{\tilde{x}} = \sum_{t\in\Z^+}\underline{x}_tz^{-t},
    \ee
    \item Spatio-Temporal Fourier transform:
    \be \label{eq:spatio_temporal_xform_def}
        \undertilde{\tilde{x}} = \sum_{k\in\Z}\sum_{t\in\Z^+}x_{k,t} e^{-ik\theta}z^{-t},
    \ee
\end{itemize}
where $i$ is the imaginary unit.\par
When a system is linear, temporally and spatially-invariant\cite{bamieh2002distributed} (LTSI), its output can be written as a convolution of the input with the spatio-temporal impulse response $G_{k,t}$:
\be \label{eq:LTSI_implulse}
    \underline{\bar{y}} = \sum_{j=-\infty}^\infty\sum_{\tau=0}^\infty G_{k-j,t-\tau}u_{j,\tau}.
\ee
Taking the spatio-temporal Fourier transform of \eqref{eq:LTSI_implulse} gives
\be
    \undertilde{\tilde{y}} = \undertilde{\tilde{G}}\undertilde{\tilde{u}}.
\ee
We measure the ``gain'' of LTSI systems with the 2-norm:
\be \label{eq:norm_defs}
    \ba
        \|\bar{\underline{G}}\|_2^2 &:= \sum_{k=-\infty}^\infty\sum_{t=0}^\infty {\rm trace}\big((G_{k,t})^TG_{k,t}\big) = \\
        \|\bar{\undertilde{G}}\|_2^2 &:= \frac{1}{2\pi} \int_{\theta=0}^{2\pi}\sum_{t=0}^\infty {\rm trace}\big((G_{\theta,t})^\dagger G_{\theta,t}\big)~d\theta = \\
        \|\underline{\tilde{G}}\|_2^2 &:= \frac{1}{2\pi} \int_{z\in\partial\mathbb{D}}\sum_{k=0}^\infty {\rm trace}\big((G_{k,z})^{\dagger}G_{k,z}\big)~dz = \\
        \|\tilde{\undertilde{G}}\|_2^2 &:= \frac{1}{4\pi^2} \int_{\theta=0}^{2\pi}\int_{z\in\partial\mathbb{D}} {\rm trace}\big((G_{\theta,z})^{\dagger}\hat{G}_{\theta,z}\big)~dz~d\theta,
    \ea
\ee
where $\mathbb{D}$ is the unit disk in $\C$.\par
We say that $\tilde{\undertilde{G}}\in\mathcal{RH}_\infty$ if $\tilde{G}^{mn}_\theta$ is a causal, stable, rational transfer function for all $m,n,\theta$. We say that $\tilde{\undertilde{G}}\in z^{-1}\mathcal{RH}_\infty$ if each element of $\tilde{\undertilde{G}}$ is additionally strictly causal.

\subsection{Factored Controller Design}
\begin{figure}[h]
    \centering
    \begin{subfigure}[]
        \centering
        \includegraphics[width=0.65\linewidth]{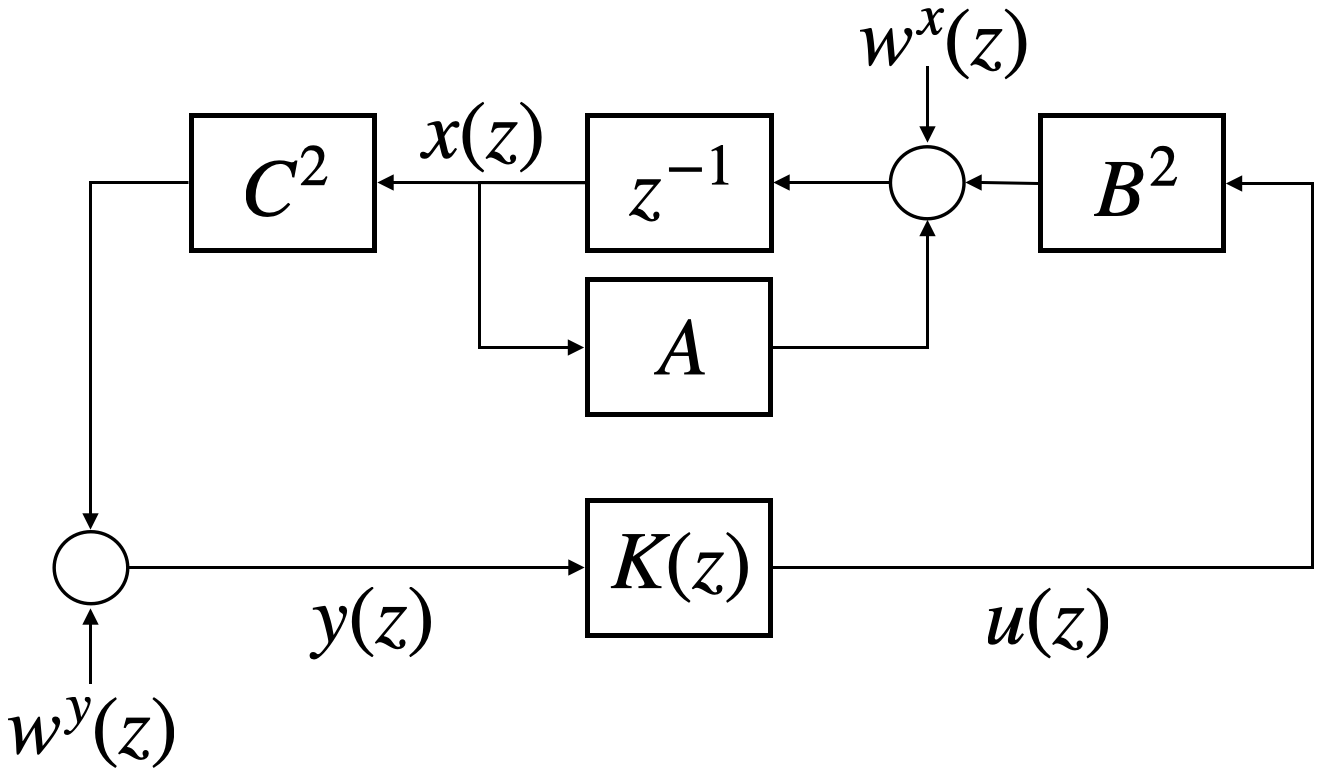}
    \end{subfigure}
    \\
    \begin{subfigure}[]
        \centering
        \includegraphics[width=0.65\linewidth]{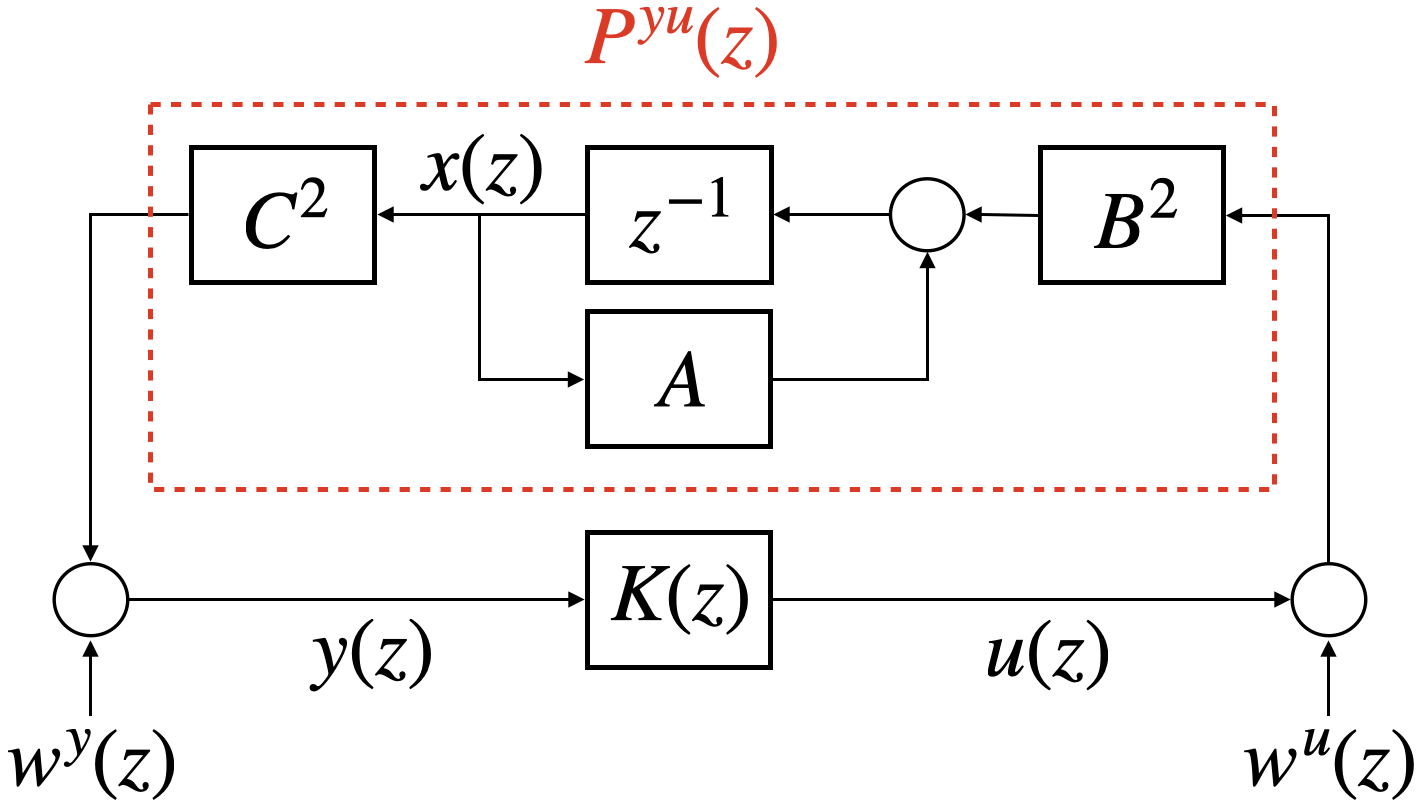}
    \end{subfigure}
    \caption{Feedback interconnection of plant and controller with external disturbances as prescribed by (a) SL and (b) IO.}
    \label{fig:internal_stability}
\end{figure}

We consider state-space models with state $x$, control input $u$, and disturbance $w$. The factored controller approaches we use, SL and IO, assume slightly different disturbance models. SL considers process disturbance, $w^x$, that enters directly into the state, i.e.
\be \label{eq:x_plus_SL}
    x_{t+1} = Ax_t + w^x_t + B^2u_t,
\ee
while IO considers disturbance, $w^u$, which enters into the control effort, i.e.
\be \label{eq:x_plus_IO}
    x_{t+1} = Ax_t + B^2\big(u_t+w^u_t\big).
\ee
Both approaches assume regulated and measurement outputs, $\zeta$ and $y$ respectively, of the following form:
\begin{subequations}
    \begin{align}
        \zeta_t &= C^1x_t + D^{12}u_t\\
        y_t &= C^2x_t + w^y_t \label{eq:measurement_eqn}.
    \end{align}
\end{subequations}
System-level and input-output parameterizations define system response maps of the closed-loop systems shown in Figure \ref{fig:internal_stability} as follows:
\begin{subequations}
    \begin{align}
        \lba{c}\tilde{x}\\\tilde{u}\ear &= \lba{cc}\tilde{R}&\tilde{N}\\\tilde{M}&\tilde{L}\ear\lba{c}\tilde{w}^x\\\tilde{w}^y\ear \label{eq:SL_maps_def}\\
        \lba{c}\tilde{y}\\\tilde{u}+\tilde{w}^u\ear &= \lba{cc}\tilde{\Gamma}&\tilde{\Lambda}\\\tilde{\Psi}&\tilde{\Omega}\ear\lba{c}\tilde{w}^y\\\tilde{w}^u\ear \label{eq:IO_maps_def}
    \end{align}
\end{subequations}
The seminal papers on system-level\cite{wang2019system} and input-output\cite{furieri2019inputoutput} parameterizations give convex parameterizations of the set of internally stabilizing \cite[Definition. 7.1]{boyd1991linear} controllers, $\mathcal{C}_{\rm stab.}$. They are given as follows:
\begin{thm} \cite[Theorem 2]{wang2019system} \label{thm:SL_Cstab}
    The affine subspace described by
    \begin{subequations}\label{eq:SL_affine}
        \begin{align}
        \lba{cc}zI-A & -B^2\ear \lba{cc}\tilde{R} & \tilde{N} \\ \tilde{M} & \tilde{L}\ear &= \lba{cc}I & 0\ear, \label{eq:SL_affine1} \\
        \lba{cc}\tilde{R} & \tilde{N} \\ \tilde{M} & \tilde{L}\ear \lba{c}zI-A \\ -C^2\ear &= \lba{c}I \\ 0\ear, \label{eq:SL_affine2} \\
        \tilde{R},\tilde{M},\tilde{N} \in z^{-1}\mathcal{RH}_\infty,~ \tilde{L} &\in \mathcal{RH}_\infty, \label{eq:SL_sets}
        \end{align}
    \end{subequations}
    gives all closed-loop responses achievable by an internally stabilizing controller. The set of internally stabilizing controllers $\tilde{u}=\tilde{K}\tilde{y}$ for \eqref{eq:x_plus_SL},\eqref{eq:measurement_eqn} is given by $\mathcal{C}_{\rm stab.} = \{\tilde{K} = \tilde{L}-\tilde{M}(\tilde{R})^{-1}\tilde{N}~|~\tilde{R},\tilde{N},\tilde{M},\tilde{L}~{\rm satisfy}~\eqref{eq:SL_affine1}-\eqref{eq:SL_sets}\}$.
\end{thm}
\begin{thm} \cite[Theorem 1]{furieri2019inputoutput} \label{thm:IO_Cstab}
    The affine subspace described by
    \begin{subequations}\label{eq:IO_affine}
        \begin{align}
        \lba{cc}I & -\tilde{P}^{yu}\ear \lba{cc}\tilde{\Gamma} & \tilde{\Lambda} \\ \tilde{\Psi} & \tilde{\Omega}\ear &= \lba{cc}I & 0\ear \label{eq:IO_affine1}, \\
        \lba{cc}\tilde{\Gamma} & \tilde{\Lambda} \\ \tilde{\Psi} & \tilde{\Omega}\ear \lba{c}-\tilde{P}^{yu} \\ I\ear &= \lba{c}0 \\ I\ear, \label{eq:IO_affine2} \\
        \tilde{\Gamma},\tilde{\Lambda},\tilde{\Psi},\tilde{\Omega} &\in \mathcal{RH}_\infty, \label{eq:IO_sets}
        \end{align}
    \end{subequations}
    where $\tilde{P}^{yu}\!=\!C^2(zI\!-\!A)^{-1}B^2$, gives all closed-loop responses achievable by an internally stabilizing controller. The set of internally stabilizing controllers $\tilde{u}=\tilde{K}\tilde{y}$ for \eqref{eq:x_plus_IO},\eqref{eq:measurement_eqn} is given by $\mathcal{C}_{\rm stab.} = \{\tilde{K} = \tilde{\Psi}(\tilde{\Gamma})^{-1}~|~\tilde{\Gamma},\tilde{\Lambda},\tilde{\Psi},\tilde{\Omega}~{\rm satisfy}~\eqref{eq:IO_affine1}-\eqref{eq:IO_sets}\}$.
\end{thm}

\subsection{Structured Systems and Extent}
Here we restate two useful definitions from \cite{jensen2020gap}, which will be used to motivate the restrictions we eventually impose on the optimal controller design problem.
\begin{defn} \cite[Definition 3.1]{jensen2020gap}
    A MIMO system $\underline{\tilde{G}}$ is \textbf{structured} with respect to a graph $\mathcal{G}$ if $\underline{\tilde{G}}^{ij}=0$ whenever $\mathcal{A}^{ij}(\mathcal{G})=0$, where $\mathcal{A}(G)$ denotes the adjacency matrix of $\mathcal{G}$.
\end{defn}
\begin{defn} \cite[Definition 3.2]{jensen2020gap}
    A realization $\lba{c|c}A&B\\\hline C&D\ear$ is a \textbf{structured realization} with respect to $\mathcal{G}$ if $A,B,C$ and $D$ are structured with respect to $\mathcal{G}$. When such a realization of the transfer function matrix $\underline{\tilde{G}}$ exists, we say $\underline{\tilde{G}}$ is structured-realizable with respect to $\mathcal{G}$.
\end{defn}
\cite[Definition 4.1]{jensen2020gap} provides a definition of an \textit{SL-structured} controller which is only applicable to systems with state feedback. The following definition is a natural extension of \cite[Definition 4.1]{jensen2020gap} to systems with output feedback.
\begin{defn}
    An LTI system $\underline{\tilde{K}}$ is an \textbf{SL-structured} controller with respect to $\mathcal{G}$ for plant realization $\underline{\tilde{P}}^{yu}=\lba{c|c}A&B^2\\\hline C^2&0\ear$ if the corresponding closed-loop transfer matrices, $\underline{\tilde{R}},\underline{\tilde{M}},\underline{\tilde{N}},\underline{\tilde{L}}$, are structured with respect to $\mathcal{G}$.
\end{defn}
We also state an analogous definition for the input-output parameterization.
\begin{defn}
    An LTI system $\underline{\tilde{K}}$ is an \textbf{IO-structured} controller with respect to $\mathcal{G}$ for plant $\underline{\tilde{P}}^{yu}$ if the corresponding closed-loop maps, $\underline{\tilde{\Gamma}}$, $\underline{\tilde{\Lambda}}$, $\underline{\tilde{\Psi}}$ and $\underline{\tilde{\Omega}}$, are structured with respect to $\mathcal{G}$.
\end{defn}
When $\underline{\tilde{K}}$ is SL-structured or IO-structured with respect to $\mathcal{G}$ for $\underline{\tilde{P}}^{yu}$ we say that the closed-loop is structured with respect to $\mathcal{G}$.\par
In the context of an infinite string of subsystems, a particular structure of interest is the $E$-nearest neighbors graph $\mathcal{G}^E_{\rm n.n.}$, whose adjacency matrix, which is Toeplitz, is given by
\be \label{eq:nearest_neighbors_graph}
    \mathcal{A}^{ij}(\mathcal{G}^E_{\rm n.n.}) = \begin{cases}
        1 & |i-j| \leq E\\
        0 & {\rm otherwise}.
    \end{cases}
\ee
We use $\mathcal{G}^E_{\rm n.n.}$ to provide definitions for \textit{extent} and \textit{communication extent}:
\begin{defn}
    An LTSI system $\underline{\tilde{G}}$ has \textbf{extent} $E$ if it is structured with respect to $\mathcal{G}^E_{\rm n.n.}$. In this case we say ${\rm ext}(\underline{\tilde{G}})=E$.
\end{defn}
\begin{defn}
    An LTSI system $\underline{\tilde{G}}$ has \textbf{communication extent} $E$ if it is structured-realizable with respect to $\mathcal{G}^E_{\rm n.n.}$. 
\end{defn}
When we consider mappings in the spatial index domain with extent $E<\infty$ and input nonzero only at $k=0$, we use subscript ($_{-E:E}$) to indicate that the mapping is being represented as a ${2E+1}\times1$ vector whose elements correspond to each spatial index $k$ with nonzero output, i.e.
\be
    \tilde{r}_{-E:E} := \lba{c}\tilde{r}_{-E}\\\vdots\\\tilde{r}_0\\\vdots\\\tilde{r}_E\ear.
\ee
Similarly, we use the subscript $(_{E_{\rm out}\leftarrow E_{\rm in}})$ on finite extent mappings when the input is nonzero only at spatial locations $-E_{\rm in}\leq k \leq E_{\rm in}$ and the output is nonzero only at spatial locations $-E_{\rm out}\leq k \leq E_{\rm out}$. For example, $\tilde{V}_{E_{\rm out}\leftarrow E_{\rm in}}$ is a $(2E_{\rm out}+1)\times(2E_{\rm in}+1)$ matrix of transfer functions.

\section{Problem Statement}
In this section we motivate the optimal controller design problem that we consider for the remainder of the paper.\par
This work aims to elucidate the complications that arise for a more general class of plants than considered in \cite{jensen2018locality}, namely LTSI systems composed of higher-order subsystems with output feedback. For simplicity of exposition, we consider optimal controller design for the following plant composed of second-order subsystems:
\be \label{eq:plant_idx}
    \ba
        x^1_{k,t+1} &= \beta x^1_{k,t} + x^2_{k,t}\\
        x^2_{k,t+1} &= \alpha\big(x^2_{k,t}+\kappa(x^2_{k-1,t}+x^2_{k+1,t})\big) + w^u_{k,t} + u_{k,t}\\
        \zeta_{k,t} &= \lba{ccc}x^1_{k,t}&x^2_{k,t}&u_{k,t}\ear^T\\
        y_{k,t} &= x^2_{k,t} + w^y_{k,t}.
    \ea
\ee
This plant is a generalization of the one used in \cite{jensen2018locality}. The dynamics of the second state, $x^2$, are identical to the dynamics of the single state in \cite[(4)]{jensen2018locality}. The parameters $\alpha$ and $\beta$ determine growth/decay rates of the two states while the parameter $\kappa$ determines coupling strength between subsystems.\par
In addition to its similarity to the plant considered in \cite{jensen2018locality}, the plant \eqref{eq:plant_idx} is composed of second order subsystems with nearest neighbor interactions, similarly to the spatially discretized 1-D wave equation and certain vehicle platoon models \cite[(47)]{zheng2022numerical}. For simplicity of exposition, we perform analysis on \eqref{eq:plant_idx}, but the same process could be applied to discrete wave or vehicle platoon models.\par
The plant \eqref{eq:plant_idx} can equivalently be written as
\be \label{eq:plant_freq}
    \ba
        \lba{c}\undertilde{x}^1_{t+1}\\\undertilde{x}^2_{t+1}\\\hline\undertilde{\zeta}^{x_1}_{t}\\\undertilde{\zeta}^{x_2}_{t}\\\undertilde{\zeta}^u_{t}\\\hline\undertilde{y}_{t}\ear &= \lba{cc|cc|c}\beta&1&0&0&0\\0&\undertilde{\sigma}&0&1&1\\\hline 1&0&0&0&0\\0&1&0&0&0\\0&0&0&0&1\\\hline 1&0&1&0&0\ear \lba{c}\undertilde{x}^1_{t}\\\undertilde{x}^2_{t}\\\hline\undertilde{w}^y_{t}\\\undertilde{w}^u_{t}\\\hline\undertilde{u}_{t}\ear\\
        & =: \lba{ccc}\undertilde{A}&B^1&B^2\\C^1&0&D^{12}\\C^2&D^{21}&0\ear\lba{c}\undertilde{x}_{t}\\\undertilde{w}_{t}\\\undertilde{u}_{t}\ear,
    \ea
\ee
where $\undertilde{\sigma} := \alpha(1+2\kappa\cos\theta)$.\par
The standard problem of optimal disturbance rejection is to design a controller $\tilde{\undertilde{u}} = \tilde{\undertilde{K}}\tilde{\undertilde{y}}$ that minimizes the $\mathcal{H}_2$ norm of the closed-loop map, $\mathcal{F}(\tilde{P},\tilde{K})$, from $\tilde{w}$ to $\tilde{\zeta}$, i.e.
\begin{subequations}\label{eq:OCP_unconstrained}
    \begin{align}
        &\inf_{\tilde{K}} ~~\|\mathcal{F}(\tilde{P},\tilde{K})\|_2 \label{eq:OCP_uncon_obj_idx}\\
        &{\rm s.t.}~~\tilde{K}\in\mathcal{C}_{\rm stab.} \label{eq:OCP_uncon_Cstab_idx}
    \end{align}
\end{subequations}
The problem \eqref{eq:OCP_unconstrained} cannot be passed to standard solvers in the spatial index domain because the plant $\underline{\tilde{P}}$ described by \eqref{eq:plant_idx} is infinite dimensional. While \eqref{eq:OCP_unconstrained} could be solved in the spatial frequency domain for the plant \eqref{eq:plant_freq}, which is finite dimensional at each single value of $\theta$, there is no guarantee that $\underline{\tilde{K}}$ recovered via inverse spatial transformation of $\tilde{\undertilde{K}}$ will have a realization with finite communication extent, making implementation difficult if not impossible. Ideally, we would solve \eqref{eq:OCP_unconstrained} subject to the additional constraint that $\underline{\tilde{K}}$ has finite communication extent, but there are no known convex formulations of the structured-realizability constraint. One approach to guarantee a controller with finite communication extent would be to impose that $\underline{\tilde{K}}$ has finite extent, but this constraint is not quadratically invariant under the plant, so the corresponding optimization problem would not be convex. \par
To achieve a controller with finite communication extent, we can instead impose that the closed-loop mappings have finite extent, as \cite{jensen2020gap} gives a constructive procedure for finding a structured realization when the closed-loop responses are structured. Therefore, we can guarantee recovery of a controller with finite communication extent by solving the following problem:\footnote{For SL, \eqref{eq:OCP_con_extent} is $\underline{\tilde{R}}$, $\underline{\tilde{M}}$, $\underline{\tilde{N}}$ and $\underline{\tilde{N}}$ have finite extent. For IO, it is $\underline{\tilde{\Gamma}}$, $\underline{\tilde{\Lambda}}$, $\underline{\tilde{\Psi}}$ and $\underline{\tilde{\Omega}}$ have finite extent.}
\begin{subequations}\label{eq:OCP_constrained}
    \begin{align}
        &\inf_{\tilde{K}} ~~\|\mathcal{F}(\tilde{P},\tilde{K})\|_2 \label{eq:OCP_con_obj}\\
        &{\rm s.t.}~~\tilde{K}\in\mathcal{C}_{\rm stab.} \label{eq:OCP_con_Cstab}\\
        &~~~~~~\text{closed-loop responses have finite extent} \label{eq:OCP_con_extent}
    \end{align}
\end{subequations}
In the following section, we utilize SL and IO to construct convex reformulations of \eqref{eq:OCP_constrained} in terms of the closed-loop responses defined in \eqref{eq:SL_maps_def} and \eqref{eq:IO_maps_def} respectively.

\section{Convex Optimal Control Problem Formulations}
In this section we state our main result: formulations of \eqref{eq:OCP_constrained} via SL and IO which are unconstrained, convex, and have a finite number of transfer function decision variables.
\begin{thm} \label{thm:OCP_SL}
    The optimization problem \eqref{eq:OCP_constrained} with constraint \eqref{eq:OCP_con_extent} as ${\rm ext}(\underline{\tilde{R}})<\infty$, ${\rm ext}(\underline{\tilde{M}})<\infty$, ${\rm ext}(\underline{\tilde{N}})<\infty$, ${\rm ext}(\underline{\tilde{L}})<\infty$ is equivalent to the following unconstrained optimization problem:
    \be \label{eq:OCP_SL}
        \inf_{\tilde{f}_{-E:E}\in\mathcal{RH}_\infty}~~\big\|\tilde{V}^{\rm SL}_{E+3\leftarrow E}\tilde{f}_{-E:E}+\tilde{h}^{\rm SL}_{-(E+3):E+3}\big\|_2^2,
    \ee
    where $\tilde{f}_{-E:E}\in\mathcal{RH}_\infty^{2E+1\times1}$ and $\tilde{V}^{\rm SL}_{E+3\leftarrow E}$ and $\tilde{h}^{\rm SL}_{-(E+3):E+3}$ are given in \eqref{eq:V_SL} and \eqref{eq:H_SL} in the proof.\footnote{The finite bounds on the extent of the SL mappings are piecewise-linear functions of $E$ which are derived in the Appendix and given in \eqref{eq:SL_extent_bounds}.}
\end{thm}
The proof of Theorem \ref{thm:OCP_SL}, which relies on three lemmas, is given in the next section.
\begin{thm}\label{thm:OCP_IO}
    The optimization problem \eqref{eq:OCP_constrained} with constraint \eqref{eq:OCP_con_extent} as ${\rm ext}(\underline{\tilde{\Gamma}})<\infty$, ${\rm ext}(\underline{\tilde{\Lambda}})<\infty$, ${\rm ext}(\underline{\tilde{\Psi}})<\infty$, ${\rm ext}(\underline{\tilde{\Omega}})<\infty$ is equivalent to the following unconstrained optimization problem:
    \be \label{eq:OCP_IO}
        \inf_{\tilde{f}_{-E:E}\in\mathcal{RH}_\infty}~~\big\|\tilde{V}^{\rm IO}_{E+3\leftarrow E}\tilde{f}_{-E:E}+\tilde{h}^{\rm IO}_{-(E+3):E+3}\big\|_2^2,
    \ee
    where $\tilde{f}_{-E:E}\in\mathcal{RH}_\infty^{2E+1\times1}$ and $\tilde{V}^{\rm IO}_{E+3\leftarrow E}$ and $\tilde{h}^{\rm IO}_{-(E+3):E+3}$ are not given directly (for concision) but are constructed in the proof analogously to $\tilde{V}^{\rm SL}_{E+3\leftarrow E}$ and $\tilde{h}^{\rm SL}_{-(E+3):E+3}$.\footnote{The finite bounds on the extent of the IO mappings are given in \eqref{eq:IO_extent_bounds}.}
\end{thm}
The proof of Theorem \ref{thm:OCP_IO} also relies on three lemmas and is given in the next section.

\section{Proofs of Main Results}
\subsection{SL Reformulation}
We first state and prove the three lemmas necessary for the SL reformulation of \eqref{eq:OCP_constrained}, Theorem \ref{thm:OCP_SL}.
\begin{lem} \label{lem:SL_Cstab_freq}
    The affine subspace described by
    \begin{subequations}\label{eq:SL_affine_freq}
        \begin{align}
        \lba{cc}zI-\undertilde{A} & -B^2\ear \lba{cc}\undertilde{\tilde{R}} & \undertilde{\tilde{N}}\\ \undertilde{\tilde{M}} & \undertilde{\tilde{L}}\ear &= \lba{cc}I & 0\ear, \label{eq:SL_affine1_freq} \\
        \lba{cc}\undertilde{\tilde{R}} & \undertilde{\tilde{N}} \\ \undertilde{\tilde{M}} & \undertilde{\tilde{L}}\ear \lba{c}zI-\undertilde{A} \\ -C^2\ear &= \lba{c}I \\ 0\ear, \label{eq:SL_affine2_freq} \\
        \undertilde{\tilde{R}},\undertilde{\tilde{M}},\undertilde{\tilde{N}} \in z^{-1}\mathcal{RH}_\infty,~ \undertilde{\tilde{L}} &\in \mathcal{RH}_\infty, \label{eq:SL_sets_freq}
        \end{align}
    \end{subequations}
    gives all closed-loop responses achievable by a controller $\undertilde{\tilde{K}} = \undertilde{\tilde{L}} - \undertilde{\tilde{M}}(\undertilde{\tilde{R}})^{-1}\undertilde{\tilde{N}}$ that internally stabilizes \eqref{eq:plant_freq}.
\end{lem}
\begin{proof}
    This is a straightforward application of \cite[Theorem 1]{bamieh2002distributed} to Theorem \ref{thm:SL_Cstab}.
\end{proof}
\begin{lem} \label{lem:SL_affine}
    For any controller which internally stabilizes \eqref{eq:plant_freq}, all scalar elements of the closed-loop maps $\undertilde{\tilde{R}},\undertilde{\tilde{M}},\undertilde{\tilde{N}},\undertilde{\tilde{L}}$ are affine in $\undertilde{\tilde{r}}^{12}$.
    Furthermore, for an internally stabilizing controller, if $\underline{\tilde{r}}^{12}$ has extent $E+1$, then each scalar element of $\underline{\tilde{R}}$, $\underline{\tilde{M}}$, $\underline{\tilde{N}}$ and $\underline{\tilde{L}}$ has extent at most $E+3$. Equivalently, for each $q\in\{r^{11},r^{12},r^{21},r^{22},m^1,m^2,n^1,n^2,\ell\}$, there exist $\tilde{V}^q_{E_q\leftarrow E+1}$ and $\tilde{h}^q_{-E_q:E_q}$ with $E_q\leq E+3$ such that
    \be \label{eq:SL_general_map_r22}
        \tilde{q}_{-E_q:E_q} = \tilde{V}^q_{E_q\leftarrow E+1}\tilde{r}^{12}_{-(E+1):E+1} + \tilde{h}^q_{-E_q:E_q}.
    \ee
\end{lem}
\begin{proof}
    Straightforward algebra allows \eqref{eq:SL_affine1_freq} and \eqref{eq:SL_affine2_freq} to be rewritten as
    \begin{subequations} \label{eq:SL_maps_of_r22}
        \begin{align}
            \undertilde{\tilde{r}}^{11} &= (z-\undertilde{\sigma})\undertilde{\tilde{r}}^{12},\label{eq:SL_r11}\\
            \undertilde{\tilde{r}}^{21} &= \undertilde{\tilde{m}}^2 = \undertilde{\tilde{n}}^1 = (z-\undertilde{\sigma})(z-\beta)\undertilde{\tilde{r}}^{12} - 1,\label{eq:SL_r21_m2_n1}\\
            \undertilde{\tilde{r}}^{22} &= (z-\beta)\undertilde{\tilde{r}}^{12},\label{eq:SL_r22}\\
            \undertilde{\tilde{m}}^{1} &= (z-\undertilde{\sigma})^2(z-\beta)\undertilde{\tilde{r}}^{12} - (z-\undertilde{\sigma}),\label{eq:SL_m1}\\
            \undertilde{\tilde{n}}^{2} &= (z-\undertilde{\sigma})(z-\beta)^2\undertilde{\tilde{r}}^{12} - (z-\beta),\label{eq:SL_n2}\\
            \undertilde{\tilde{\ell}} &= (z-\undertilde{\sigma})^2(z-\beta)^2\undertilde{\tilde{r}}^{12} - (z-\beta)(z-\undertilde{\sigma}).\label{eq:SL_l}
        \end{align}
    \end{subequations}
    Clearly each $\undertilde{\tilde{q}}$ is explicitly affine in $\undertilde{\tilde{r}}^{12}$, so after inverse spatial transformation, each $\tilde{q}_k$ will be affine in $\tilde{r}^{12}_j$ for $j$ in some set of spatial locations. The operator $\undertilde{\sigma}$ adds one to the extent of a mapping, i.e. if $\undertilde{\tilde{q}} = (\undertilde{\sigma})^j\undertilde{\tilde{r}}^{12}$ and ${\rm ext}(\underline{\tilde{r}}^{12})=E$, then ${\rm ext}(\underline{\tilde{q}})=E+j$. The operator $\undertilde{\sigma}$ is applied to $\undertilde{\tilde{r}}^{12}$ at most twice in \eqref{eq:SL_maps_of_r22}, so the extent of any closed-loop map will be at most two more than the extent of $\underline{\tilde{r}}^{12}$.
\end{proof}
\begin{lem} \label{lem:SL_tilde}
    A controller internally stabilizes \eqref{eq:plant_freq} if and only if the corresponding closed-loop maps satisfy \eqref{eq:SL_affine1_freq} and \eqref{eq:SL_affine2_freq} and
    \be \label{eq:SL_R22_expand}
        \tilde{\undertilde{r}}^{12} = z^{-2} + z^{-3}(\undertilde{\sigma}+\beta) + z^{-4}\undertilde{\tilde{f}},
    \ee
    where $\undertilde{\tilde{f}}\in\mathcal{RH}_\infty$.
\end{lem}
\begin{proof}
    As shown in Lemma \ref{lem:SL_affine}, \eqref{eq:SL_affine1_freq} and \eqref{eq:SL_affine2_freq} are equivalent to \eqref{eq:SL_maps_of_r22}. By Lemma \ref{lem:SL_Cstab_freq}, for internal stability, \eqref{eq:SL_sets_freq} must also be satisfied. It's straightforward to see in \eqref{eq:SL_maps_of_r22} that $\tilde{\undertilde{n}}^2$ and $\tilde{\undertilde{\ell}}$ have the highest relative degrees compared to the sets that \eqref{eq:SL_sets_freq} prescribes they be elements of. Therefore when \eqref{eq:SL_affine1_freq} and \eqref{eq:SL_affine2_freq} are satisfied, \eqref{eq:SL_sets_freq} is equivalent to $\undertilde{\tilde{\ell}}\in z^{-1}\mathcal{RH}_\infty$. Without loss of generality, $\undertilde{\tilde{r}}^{12}\in z^{-1}\mathcal{RH}_\infty$ can be decomposed as
    \be \label{eq:SL_R22_generalform}
        \undertilde{\tilde{r}}^{12} = \undertilde{\rho}^1 z^{-1} + \undertilde{\rho}^2 z^{-2} + \undertilde{\rho}^3 z^{-3} + z^{-4}\undertilde{\tilde{f}}
    \ee
    with $\undertilde{\rho}^1$ and $\undertilde{\rho}^2$ static and $\undertilde{\tilde{f}}\in\mathcal{RH}_\infty$. Then calculating $\undertilde{\tilde{\ell}}$ in terms of $\undertilde{\tilde{f}}$ from \eqref{eq:SL_l} and \eqref{eq:SL_R22_generalform},
    \be
        \ba
            \undertilde{\tilde{\ell}} &= z^{-4}\Big(\undertilde{\rho}^1 z^7 + \big(\undertilde{\rho}^2-2(\undertilde{\sigma}\!+\!\beta)\undertilde{\rho}^1-1\big)z^6 \\
            &+ \big(\undertilde{\rho}^3 - 2(\undertilde{\sigma}\!+\!\beta)\undertilde{\rho}^2 + (\undertilde{\sigma}^2\!+\!\beta^2\!+\!4\undertilde{\sigma}\beta)\undertilde{\rho}^1 + \undertilde{\sigma} + \beta\big)z^5 \\
            &+ \mathcal{O}(z^4)\Big).
        \ea
    \ee
    For $\undertilde{\tilde{\ell}}$ to be causal, its necessary and sufficient that $\undertilde{\rho}^1=0$, $\undertilde{\rho}^2=1$ and $\undertilde{\rho}^3=\undertilde{\sigma}+\beta$. Then \eqref{eq:SL_R22_generalform} simplifies to \eqref{eq:SL_R22_expand}.
\end{proof}
We now apply these three lemmas to prove Theorem \ref{thm:OCP_SL}.
\begin{proof}[Proof of Theorem \ref{thm:OCP_SL}]
    The cost function of \eqref{eq:OCP_constrained} is the $\mathcal{H}_2$ norm of $\mathcal{F}(\tilde{P},\tilde{K})$, the closed-loop mapping from $\tilde{w}$ to $\tilde{\zeta}$. In terms of the SL closed-loop mappings,
    \be
        \ba
            \mathcal{F}(\undertilde{\tilde{P}},\undertilde{\tilde{K}}) &= \lba{cc}C^1&D^{12}\ear\lba{cc}\undertilde{\tilde{R}}&\undertilde{\tilde{N}}\\\undertilde{\tilde{M}}&\undertilde{\tilde{L}}\ear\lba{c}B^1\\D^{21}\ear\\
            &= \lba{cc}\undertilde{\tilde{n}}^1&\undertilde{\tilde{r}}^{12}\\\undertilde{\tilde{n}}^2&\undertilde{\tilde{r}}^{22}\\\undertilde{\tilde{\ell}}&\undertilde{\tilde{m}}^2\ear.
        \ea
    \ee
    We can then apply properties of the $\mathcal{H}_2$ norm to rewrite the cost function as
    \be \label{eq:SL_cost_rearrange}
        \Big\| \lba{cc}\undertilde{\tilde{n}}^1&\undertilde{\tilde{r}}^{12}\\\undertilde{\tilde{n}}^2&\undertilde{\tilde{r}}^{22}\\\undertilde{\tilde{\ell}}&\undertilde{\tilde{m}}^2\ear \Big\|_2 = \Big\| \lba{c}\undertilde{\tilde{n}}^1\\\undertilde{\tilde{r}}^{12}\\\undertilde{\tilde{n}}^2\\\undertilde{\tilde{r}}^{22}\\\undertilde{\tilde{\ell}}\\\undertilde{\tilde{m}}^2\ear \Big\|_2 = \Big\| \lba{c}\underline{\tilde{n}}^1\\\underline{\tilde{r}}^{12}\\\underline{\tilde{n}}^2\\\underline{\tilde{r}}^{22}\\\underline{\tilde{\ell}}\\\underline{\tilde{m}}^2\ear \Big\|_2.
    \ee
    Inverse transforming the affine maps given in Lemma \ref{lem:SL_affine}, \eqref{eq:SL_maps_of_r22}, which appear in the cost function \eqref{eq:SL_cost_rearrange}, for $\underline{\tilde{r}}^{12}$ with extent $E+1$, we find
    \begingroup \small
    \begin{subequations} \label{eq:SL_affine_in_R22}
        \begin{align}
            \tilde{n}^1_{-(E+2):E+2} &= \tilde{V}^{n^1}_{E+2\leftarrow E+1}\tilde{r}^{12}_{-(E+1):E+1} + \tilde{h}^{n^1}_{-(E+2):E+2},\\
            \tilde{r}^{12}_{-(E+1):E+1} &= \tilde{V}^{r^{12}}_{E+1\leftarrow E+1}\tilde{r}^{12}_{-(E+1):E+1} + \tilde{h}^{r^{12}}_{-(E+1):E+1},\\
            \tilde{n}^2_{-(E+2):E+2} &= \tilde{V}^{n^2}_{E+2\leftarrow E+1}\tilde{r}^{12}_{-(E+1):E+1} + \tilde{h}^{n^2}_{-(E+2):E+2},\\
            \tilde{r}^{22}_{-(E+1):E+1} &= \tilde{V}^{r^{22}}_{E+1\leftarrow E+1}\tilde{r}^{12}_{-(E+1):E+1} + \tilde{h}^{r^{22}}_{-(E+1):E+1},\\
            \tilde{\ell}_{-(E+3):E+3} &= \tilde{V}^{\ell}_{E+3\leftarrow E+1}\tilde{r}^{12}_{-(E+1):E+1} + \tilde{h}^{\ell}_{-(E+3):E+3},\\
            \tilde{m}^2_{-(E+2):E+2} &= \tilde{V}^{m^2}_{E+2\leftarrow E+1}\hat{r}^{12}_{-(E+1):E+1} + \tilde{h}^{m^2}_{-(E+2):E+2},
        \end{align}
    \end{subequations}
    \endgroup
    where for each $q\in\{n^1,r^{12},n^2,r^{22},\ell,m^2\}$, $\tilde{V}^q_{E^q\leftarrow E+1}$ and $\tilde{h}^q_{-E^q:E^q}$ are given by \eqref{eq:SL_V_H} in the Appendix. 
    Inverse transforming the affine map given in Lemma \ref{lem:SL_tilde}, \eqref{eq:SL_R22_expand}, we find
    \be \label{eq:r22_of_rtil}
        \tilde{r}^{12}_{-(E+1):E+1} = \tilde{S}_{E+1\leftarrow E}\tilde{f}_{-E:E} + \tilde{g}_{-(E+1):E+1},
    \ee
    where $\tilde{S}_{E+1\leftarrow E}$ and $\tilde{g}_{-(E+1):E+1}$ are given by
    \begin{subequations}
        \begin{align}
            \tilde{S}_{E+1\leftarrow E} &= z^{-4}\lba{c}0^{1\times E}\\I^{E\times E}\\0^{1\times E}\ear \label{eq:Vtilde_SL},\\
            \tilde{g}_{-(E+1):E+1} &= z^{-3}\lba{c}0^{E\times1}\\\alpha\kappa\\z+\alpha+\beta\\\alpha\kappa\\0^{E\times1}\ear.\label{eq:Htilde_SL}
        \end{align}
    \end{subequations}
    Then from \eqref{eq:SL_affine_in_R22} and \eqref{eq:r22_of_rtil}, for each $q\in\{r^{12},r^{22},m^2,n^1,n^2,\ell\}$,
    \be
        \ba
            \tilde{q}_{-E^q:E^q} =~&  \tilde{V}^q_{E^q\leftarrow E+1}\tilde{S}_{E+1\leftarrow E}\tilde{f}_{-E:E} \\ &~~~~+ \tilde{V}^q_{E^q\leftarrow E+1}\tilde{g}_{-(E+1):E+1} + \tilde{h}^q_{-E^q:E^q},
        \ea
    \ee
    and stacking these for each map $q\in\{n^1,r^{12},n^2,r^{22},\ell,m^2\}$ as in \eqref{eq:SL_cost_rearrange}, we arrive at the cost function of \eqref{eq:OCP_SL}, where $\tilde{V}^{\rm SL}_{E+3\leftarrow E}$ is given by
    \be
        \tilde{V}^{\rm SL}_{E+3\leftarrow E} = \left[\!\!\begin{array}{c}\tilde{V}^{n^1}_{E+2\leftarrow E+1}\\\tilde{V}^{r^{12}}_{E+1\leftarrow E+1}\\\tilde{V}^{n^2}_{E+2\leftarrow E+1}\\\tilde{V}^{r^{22}}_{E+1\leftarrow E+1}\\\tilde{V}^{\ell}_{E+3\leftarrow E+1}\\\tilde{V}^{m^2}_{E+2\leftarrow E+1}\end{array}\!\!\right]\tilde{S}_{E+1\leftarrow E} \label{eq:V_SL}
    \ee
    and $\tilde{h}^{\rm SL}_{-(E+3):E+3}$ is given by
    {\small
    \be
        \tilde{h}^{\rm SL}_{-(E+3):E+3} = \left[\!\!\begin{array}{c}\tilde{V}^{n^1}_{E+2\leftarrow E+1}\\\tilde{V}^{r^{12}}_{E+1\leftarrow E+1}\\\tilde{V}^{n^2}_{E+2\leftarrow E+1}\\\tilde{V}^{r^{22}}_{E+1\leftarrow E+1}\\\tilde{V}^{\ell}_{E+3\leftarrow E+1}\\\tilde{V}^{m^2}_{E+2\leftarrow E+1}\end{array}\!\!\right] \tilde{g}_{-(E+1):E+1} + \left[\!\!\begin{array}{c}\tilde{h}^{n^1}_{-(E+2):E+2}\\\tilde{h}^{r^{12}}_{-(E+1):E+1}\\\tilde{h}^{n^2}_{-(E+2):E+2}\\\tilde{h}^{r^{22}}_{-(E+1):E+1}\\\tilde{h}^{\ell}_{-(E+3):E+3}\\\tilde{h}^{m^2}_{-(E+2):E+2}\end{array}\!\!\right]. \label{eq:H_SL}
    \ee
    }
\end{proof}

\subsection{IO Reformulation}
Next we state and prove the three lemmas necessary to prove Theorem \ref{thm:OCP_IO}. These lemmas are directly analogous to lemmas \ref{lem:SL_Cstab_freq}, \ref{lem:SL_affine} and \ref{lem:SL_tilde}, and the proof structures are similar, so we use more concision here than in the previous subsection.
\begin{lem} \label{lem:IO_Cstab_freq}
    The affine subspace described by
    \begin{subequations}\label{eq:IO_affine_freq}
        \begin{align}
        \lba{cc}I & -\undertilde{\tilde{P}}^{yu}\ear \lba{cc}\undertilde{\tilde{\Gamma}} & \undertilde{\tilde{\Lambda}} \\ \undertilde{\tilde{\Psi}} & \undertilde{\tilde{\Omega}}\ear &= \lba{cc}I & 0\ear, \label{eq:IO_affine1_freq} \\
        \lba{cc}\undertilde{\tilde{\Gamma}} & \undertilde{\tilde{\Lambda}} \\ \undertilde{\tilde{\Psi}} & \undertilde{\tilde{\Omega}}\ear \lba{c}-\undertilde{\tilde{P}}^{yu} \\ I\ear &= \lba{c}0 \\ I\ear \label{eq:IO_affine2_freq}, \\
        \undertilde{\tilde{\Gamma}},\undertilde{\tilde{\Lambda}},\undertilde{\tilde{\Psi}},\undertilde{\tilde{\Omega}} &\in \mathcal{RH}_\infty, \label{eq:IO_sets_freq}
        \end{align}
    \end{subequations}
    gives all closed-loop responses achievable by a controller $\undertilde{\tilde{K}} = \undertilde{\tilde{\Psi}}(\undertilde{\tilde{\Gamma}})^{-1}$ that internally stabilizes \eqref{eq:plant_freq}.
\end{lem}
\begin{proof}
    This is a straightforward application of \cite[Theorem 1]{bamieh2002distributed} to Theorem \ref{thm:IO_Cstab}.
\end{proof}
\begin{lem} \label{lem:IO_affine}
    For any controller which internally stabilizes \eqref{eq:plant_freq}, the scalar maps $\undertilde{\tilde{\Gamma}}, \undertilde{\tilde{\Lambda}},\undertilde{\tilde{\Psi}},\undertilde{\tilde{\Omega}}$ are affine in $\undertilde{\tilde{\lambda}}$.
    Furthermore, for an internally stabilizing controller if $\underline{\tilde{\lambda}}$ has extent $E+1$, then each of the scalar maps $\undertilde{\tilde{\gamma}}, \undertilde{\tilde{\lambda}},\undertilde{\tilde{\psi}},\undertilde{\tilde{\omega}}$ has extent at most $E+3$. Equivalently, for each $q\in\{\gamma,\lambda,\psi,\omega\}$, there exist $V^q_{E_q\leftarrow E+1}$ and $h^q_{-E_q:E_q}$ with $E_q\leq E+3$ such that
    \be \label{eq:IO_general_map_lambda}
        \tilde{q}_{-E_q:E_q} = \tilde{V}^q_{E_q\leftarrow E+1}\tilde{\lambda}_{-(E+1):E+1} + \tilde{h}^q_{-E_q:E_q}.
    \ee
\end{lem}
\begin{proof}
    Straightforward algebra allows \eqref{eq:IO_affine2_freq} and \eqref{eq:IO_affine1_freq} to be rewritten as
    \begin{subequations} \label{eq:IO_maps_of_lambda}
        \begin{align}
            \undertilde{\tilde{\gamma}} &= \undertilde{\tilde{\omega}} = (z-\beta)(z-\undertilde{\sigma})\undertilde{\tilde{\lambda}}, \label{eq:IO_gamma_omega}\\
            \undertilde{\tilde{\psi}} &= (z-\beta)^2(z-\undertilde{\sigma})^2\undertilde{\tilde{\lambda}} - (z-\beta)(z-\undertilde{\sigma}). \label{eq:IO_psi}
        \end{align}
    \end{subequations}
    The two applications of the $\undertilde{\sigma}$ operator to $\undertilde{\tilde{\lambda}}$ in \eqref{eq:IO_psi} means that $\underline{\tilde{\psi}}$ will have extent two more than that of $\underline{\tilde{\lambda}}$, while $\underline{\tilde{\gamma}}$ and $\underline{\tilde{\omega}}$ will have extent only one more than that of $\underline{\tilde{\lambda}}$.
\end{proof}
\begin{lem} \label{lem:IO_tilde}
    A controller internally stabilizes \eqref{eq:plant_freq} if and only if the corresponding closed-loop maps satisfy \eqref{eq:IO_affine1_freq} and \eqref{eq:IO_affine2_freq} and
    \be \label{eq:IO_lambda_expand}
        \undertilde{\tilde{\lambda}} = z^{-2} + z^{-3}(\undertilde{\sigma}+\beta) + z^{-4}\undertilde{\tilde{f}},
    \ee
    where $\undertilde{\tilde{f}}\in\mathcal{RH}_\infty$.
\end{lem}
\begin{proof}
    The proof of this lemma is identical to the proof of Lemma \ref{lem:SL_tilde}, with $\psi$ taking the place of $\ell$ and $\lambda$ taking the place of $r^{12}$.
\end{proof}
We now use these lemmas to prove Theorem \ref{thm:OCP_IO}.
\begin{proof}[Proof of Theorem \ref{thm:OCP_IO}]
    We include the second state, $x^2$, in the regulated output $\zeta$, but do not measure it, so in order to write the cost function in terms of the closed-loop maps, we need to write the second state, $\undertilde{\tilde{x}}^2$, as affine in $\undertilde{\tilde{w}}^y$ and $\undertilde{\tilde{w}}^u$. Temporally transforming the state evolution equation, $\undertilde{x}^1_{t+1}  = \beta\undertilde{x}^1_t + \undertilde{x}^2_t$, we find $\undertilde{\tilde{x}}^2 = (z-\beta)\undertilde{\tilde{x}}^1$, which is useful because $\undertilde{\tilde{x}}^1$ can be expressed explicitly in terms of the disturbances and the closed-loop maps as $\undertilde{\tilde{x}}^1=(\undertilde{\tilde{\gamma}}-1)\undertilde{\tilde{w}}^y + \undertilde{\tilde{\lambda}}\undertilde{\tilde{w}}^u$. Then writing $\undertilde{\tilde{x}}^2$ in terms of the disturbances and closed-loop maps,
    \be \label{eq:x2_recovery}
            \undertilde{\tilde{x}}^2 = (z-\beta)\big((\undertilde{\tilde{\gamma}}-1)\undertilde{\tilde{w}}^y + \undertilde{\tilde{\lambda}}\undertilde{\tilde{w}}^u\big).
    \ee
    The map from disturbance $\undertilde{\tilde{w}}$ to regulated output $\undertilde{\tilde{\zeta}}$ is then given by
    \be
        \ba
            \mathcal{F}(\undertilde{\tilde{P}},\undertilde{\tilde{K}}) = \lba{cc} \undertilde{\tilde{\gamma}}-1&\undertilde{\tilde{\lambda}}\\(z-\beta)(\undertilde{\tilde{\gamma}}-1)&(z-\beta)\undertilde{\tilde{\lambda}}\\\undertilde{\tilde{\psi}}&\undertilde{\tilde{\omega}}-1 \ear.
        \ea
    \ee
    Applying properties of the $\mathcal{H}_2$ norm,
    \be \label{eq:IO_cost_rearrange}
        \ba
            \Big\|\! \lba{cc} \undertilde{\tilde{\gamma}}-1&\undertilde{\tilde{\lambda}}\\(z\!-\!\beta)(\undertilde{\tilde{\gamma}}\!-\!1)&(z\!-\!\beta)\undertilde{\tilde{\lambda}}\\\undertilde{\tilde{\psi}}&\undertilde{\tilde{\omega}}-1 \ear \!\Big\|_2 &= \Big\|\! \lba{cc} \undertilde{\tilde{\gamma}}-1\\\undertilde{\tilde{\lambda}}\\(z\!-\!\beta)(\undertilde{\tilde{\gamma}}\!-\!1)\\(z-\beta)\undertilde{\tilde{\lambda}}\\\undertilde{\tilde{\psi}}\\\undertilde{\tilde{\omega}}-1 \ear \!\Big\|_2 \\
            &= \Big\| \lba{cc} \underline{\tilde{\gamma}}-1\\\underline{\tilde{\lambda}}\\(z\!-\!\beta)(\underline{\tilde{\gamma}}\!-\!1)\\(z-\beta)\underline{\tilde{\lambda}}\\\underline{\tilde{\psi}}\\\underline{\tilde{\omega}}-1 \ear \!\Big\|_2.
        \ea
    \ee
    The proof of Theorem \ref{thm:OCP_SL} uses \eqref{eq:SL_cost_rearrange}, Lemma \ref{lem:SL_affine} and Lemma \ref{lem:SL_tilde} to construct a finite extent expression for $\mathcal{F}(\underline{\tilde{P}},\underline{\tilde{K}})$ which is affine in the finite extent decision variable $\tilde{f}_{-E:E}$. The same logic used to construct $\tilde{V}^{\rm SL}_{E+3\leftarrow E}$ and $\tilde{h}^{\rm SL}_{-(E+3):E+3}$ can be applied to construct $\tilde{V}^{\rm IO}_{E+3\leftarrow E}$ and $\tilde{h}^{\rm IO}_{-(E+3):E+3}$ from \eqref{eq:IO_cost_rearrange}, Lemma \ref{lem:IO_affine} and Lemma \ref{lem:IO_tilde}, giving a finite extent expression for $\mathcal{F}(\underline{\tilde{P}},\underline{\tilde{K}})$ which is affine in the finite extent decision variable $\tilde{f}_{-E:E}$.
\end{proof}

\section{Equivalence Between SL and IO}
In this section we show that even after applying constraint \eqref{eq:OCP_con_extent}, reformulation of \eqref{eq:OCP_constrained} via SL and IO result in equivalent optimization problems. We do so by showing that the locality-constrained feasible sets for the SL and IO problems are identical.\par
We first note that Lemmas \ref{lem:SL_Cstab_freq}, \ref{lem:SL_affine} and \ref{lem:SL_tilde} allow us to express each of the closed-loop SL maps for an internally stabilizing controller as an affine function of a scalar transfer function parameter $\undertilde{\tilde{f}}$. Just as Lemmas \ref{lem:SL_Cstab_freq}, \ref{lem:SL_affine} and \ref{lem:SL_tilde} give us $\undertilde{\tilde{R}}(\undertilde{\tilde{f}})$, $\undertilde{\tilde{M}}(\undertilde{\tilde{f}})$, $\undertilde{\tilde{N}}(\undertilde{\tilde{f}})$ and $\undertilde{\tilde{L}}(\undertilde{\tilde{f}})$, Lemmas \ref{lem:IO_Cstab_freq}, \ref{lem:IO_affine} and \ref{lem:IO_tilde} give us the IO maps, $\undertilde{\tilde{\Gamma}}(\undertilde{\tilde{f}})$, $\undertilde{\tilde{\Lambda}}(\undertilde{\tilde{f}})$, $\undertilde{\tilde{\Psi}}(\undertilde{\tilde{f}})$ and $\undertilde{\tilde{\Omega}}(\undertilde{\tilde{f}})$ for an internally stabilizing controller. We can then express the unconstrained feasible set as
\be
    \ba
        \mathcal{C}_{\rm stab.} &= \big\{ \undertilde{\tilde{L}}(\undertilde{\tilde{f}})-\undertilde{\tilde{M}}(\undertilde{\tilde{f}})\big(\undertilde{\tilde{R}}(\undertilde{\tilde{f}})\big)^{-1}\undertilde{\tilde{N}}(\undertilde{\tilde{f}}) ~|~ \undertilde{\tilde{f}}\in\mathcal{RH}_\infty \big\}\\
        &= \big\{ \undertilde{\tilde{\Psi}}(\undertilde{\tilde{f}})\big(\undertilde{\tilde{\Gamma}}(\undertilde{\tilde{f}})\big)^{-1} ~|~ \undertilde{\tilde{f}}\in\mathcal{RH}_\infty \}.
    \ea
\ee
We define locality-constrained feasible sets, $\mathcal{K}^{\rm SL}_E$ and $\mathcal{K}^{\rm IO}_E$, as
\begin{subequations}
    \begin{align}
        \mathcal{K}^{\rm SL}_E &:= \big\{ \undertilde{\tilde{L}}(\undertilde{\tilde{f}})-\undertilde{\tilde{M}}(\undertilde{\tilde{f}})\big(\undertilde{\tilde{R}}(\undertilde{\tilde{f}})\big)^{-1}\undertilde{\tilde{N}}(\undertilde{\tilde{f}}) ~|~ \notag\\ &~~~~~~~~~~~~~~~~~~~~~~ \undertilde{\tilde{f}}\in\mathcal{RH}_\infty,~{\rm ext}(\underline{\tilde{f}})=E \big\}, \label{eq:K_E_SL}\\
        \mathcal{K}^{\rm IO}_E &:= \big\{ \undertilde{\tilde{\Psi}}(\undertilde{\tilde{f}})\big(\undertilde{\tilde{\Gamma}}(\undertilde{\tilde{f}})\big)^{-1} ~|~\undertilde{\tilde{f}}\in\mathcal{RH}_\infty,~{\rm ext}(\underline{\tilde{f}})=E \big\}.\label{eq:K_E_IO}
    \end{align}
\end{subequations}

We now state the main result of this section: the equivalence of the optimal control problems for SL and IO.
\begin{thm}
    For any $E$, $\mathcal{K}^{\rm SL}_E=\mathcal{K}^{\rm IO}_E$.
\end{thm}
\begin{proof}
    Clearly, given \eqref{eq:K_E_SL} and \eqref{eq:K_E_IO}, it is sufficient to show that $\undertilde{\tilde{L}}(\undertilde{\tilde{f}})-\undertilde{\tilde{M}}(\undertilde{\tilde{f}})\big(\undertilde{\tilde{R}}(\undertilde{\tilde{f}})\big)^{-1}\undertilde{\tilde{N}}(\undertilde{\tilde{f}}) = \undertilde{\tilde{\Psi}}(\undertilde{\tilde{f}})\big(\undertilde{\tilde{\Gamma}}(\undertilde{\tilde{f}})\big)^{-1}$.\par
    Using \eqref{eq:SL_maps_of_r22} and \eqref{eq:SL_R22_expand} and algebraic simplification, we arrive at an explicit expression for $\undertilde{\tilde{L}}(\undertilde{\tilde{f}})-\undertilde{\tilde{M}}(\undertilde{\tilde{f}})\big(\undertilde{\tilde{R}}(\undertilde{\tilde{f}})\big)^{-1}\undertilde{\tilde{N}}(\undertilde{\tilde{f}})$. Similarly, we can arrive at an explicit expression for $\undertilde{\tilde{\Psi}}(\undertilde{\tilde{f}})\big(\undertilde{\tilde{\Gamma}}(\undertilde{\tilde{f}})\big)^{-1}$ by using \eqref{eq:IO_maps_of_lambda} and \eqref{eq:IO_lambda_expand}. These explicit expressions are identical and are given by
    \begingroup \small
    \be
        \ba
             \undertilde{\tilde{\Psi}}(\undertilde{\tilde{f}})\big(\undertilde{\tilde{\Gamma}}(\undertilde{\tilde{f}})\big)^{-1} &= \undertilde{\tilde{L}}(\undertilde{\tilde{f}})-\undertilde{\tilde{M}}(\undertilde{\tilde{f}})\big(\undertilde{\tilde{R}}(\undertilde{\tilde{f}})\big)^{-1}\undertilde{\tilde{N}}(\undertilde{\tilde{f}})\\
            &= \frac{ (z-\beta)\big(\undertilde{\tilde{f}}(z-\undertilde{\sigma}) - z\undertilde{\sigma}(\undertilde{\sigma}+\beta)\big)-z^2\beta^2 }{ z^2 + (\undertilde{\sigma}+\beta)z + \undertilde{\tilde{f}} }.
        \ea
    \ee
    \endgroup
\end{proof}
We note that the equivalence of the SL and IO optimal control problems is not generally guaranteed for all LTSI plants. \cite{zheng2021equivalence} shows that equivalence can be guaranteed when imposed constraints are quadratically invariant under the plant. This is not the case for the chosen plant, but because $B^2$ has no spatial frequency dependence, enforcing structural constraints on maps from $\underline{\tilde{w}}^x$ ends up being equivalent to enforcing the same structural constraints on maps from $\underline{\tilde{w}}^u$. Similarly, because $C^2$ has no spatial frequency dependence, enforcing structural constraints on maps to $\underline{\tilde{x}}$ ends up being equivalent to enforcing the same structural constraints on maps to $\underline{\tilde{y}}$.\par

\section{Numerical Results}
We utilize standard model-matching techniques to solve the optimal controller synthesis problem numerically with Matlab's \texttt{h2syn}. Allowing $\underline{\tilde{f}}$ to be a vector of transfer function variables of size $2E+1$, we construct $\tilde{V}_{E+3\leftarrow E}$ and $\tilde{h}_{-(E+3):E+3}$ of the appropriate sizes. We refer to the cost, $\|\mathcal{F}(\underline{\tilde{P}},\underline{\tilde{K}})\|_{2}$, of the finite extent problem as $J_E$. We numerically solve for $J_E$ for $E\in\{1,2,\dots,10\}$.\par
\begin{figure}[h]
    \centering
    \begin{subfigure}[]
        \centering
        \includegraphics[width=0.9\linewidth]{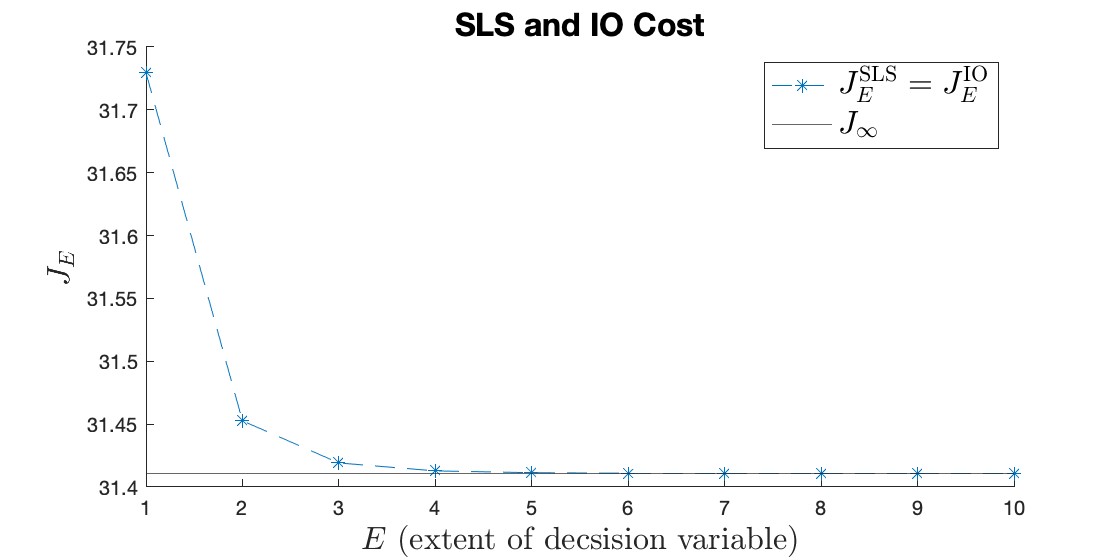}
    \end{subfigure}
    \\
    \begin{subfigure}[]
        \centering
        \includegraphics[width=0.9\linewidth]{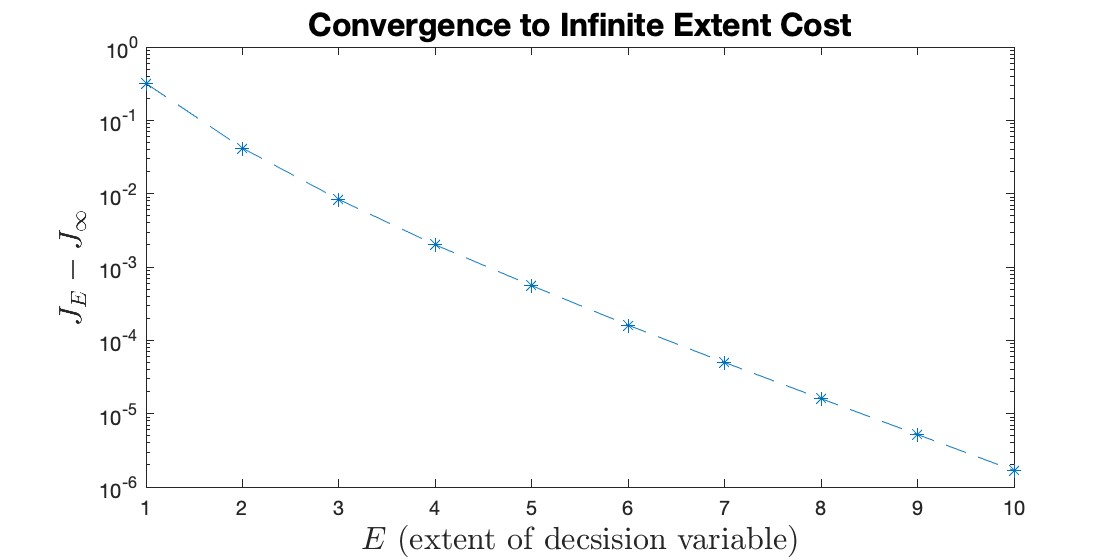}
    \end{subfigure}
    \caption{Finite extent cost comparison to infinite extent cost for $\alpha=1.5$, $\kappa=0.8$, $\beta=1$. }
    \label{fig:numeric}
\end{figure}

Figure \ref{fig:numeric} shows the convergence of the finite extent cost to $J_\infty$, the cost of the unconstrained problem \eqref{eq:OCP_unconstrained}, equivalent to allowing infinite extent closed-loop maps. $J_\infty$ is approximated by solving \eqref{eq:OCP_unconstrained} for each $\theta$ in a high-resolution discretization of $[0,2\pi)$ and numerically integrating over $\theta$ as in \eqref{eq:norm_defs}.

\section{Conclusion}
In this work, we show that the work of \cite{jensen2018locality} generalizes to a broader class of systems: namely certain second-order subsystems with output feedback. We also show that the input-output parameterization can be used just as easily in this setting, and that the problems are equivalent for the chosen plant. Our numerical example shows that increasing the closed-loop response extent beyond $E=2$ results in minute performance increases, suggesting that the communication savings introduced by using a highly local controller may outweigh the performance degradation.\par
Future work could apply these techniques to wave equation or vehicle platoon models to determine the communication-performance tradeoff for those systems. Our infinite-impulse-response techniques should be compared to finite-impulse-response synthesis with regards to performance and controller realization order. A third factored controller framework, network realization functions \cite{sabau2023network}, was not considered here but should be in the future.

\bibliography{bibliography}
\bibliographystyle{ieeetr}

\section{Appendix}

\subsection{Maps in Spatial Index}
The proof of Theorem \ref{thm:OCP_SL} makes use of $\tilde{V}^q_{E^q\leftarrow E+1}$ and $\tilde{h}^q_{-E^q:E^q}$ for $q\in\{n^1,r^{12},n^2,r^{22},\ell,m^2\}$, which we derive here, simply by taking the inverse spatial transforms of \eqref{eq:SL_maps_of_r22}.

\arraycolsep=1.3pt\def\arraystretch{0.8}
\begingroup \footnotesize
\begin{subequations} \label{eq:SL_V_H}
    \begin{align}
        \tilde{V}^{n^1}_{E+2\leftarrow E+1} &= (z\!-\!\beta)\lba{cc}-\alpha\kappa&\\z\!-\!\alpha&\ddots\\-\alpha\kappa&\ddots\\&\ddots\ear,~ \tilde{h}^{n^1}_{-(E+2):E+2} = \lba{c}0^{E+2}\\-1\\0^{E+2}\ear, \label{eq:Vh_n1}\\
        \tilde{V}^{r^{12}}_{E+1\leftarrow E+1} &= I,~~ \tilde{h}^{r^{12}}_{-(E+1):E+1} = 0^{2E+3},\label{eq:Vh_r12}\\
        \tilde{V}^{n^2}_{E+2\leftarrow E+1} &= (z\!-\!\beta)\tilde{V}^{n^1}_{E+2\leftarrow E+1},\notag\\&~~~~~~~~~~~~~~~~~ \tilde{h}^{n^2}_{-(E+2):E+2} = (z\!-\!\beta)\tilde{h}^{n^1}_{-(E+2):E+2}, \label{eq:Vh_n2}\\
        \tilde{V}^{r^{22}}_{E+1\leftarrow E+1} &= (z-\beta)I,~~ \tilde{h}^{r^{22}}_{-(E+1):E+1} = 0^{2E+3},\label{eq:Vh_r22}\\
        \tilde{V}^{\ell}_{E+1\leftarrow E+3} &= (z\!-\!\beta)^2\lba{cc}(\alpha\kappa)^2&\\-2\alpha\kappa(z\!-\!\alpha)&\ddots\\(z\!-\!\alpha)^2\!+\!2(\alpha\kappa)^2&\ddots\\-2\alpha\kappa(z\!-\!\alpha)&\ddots\\(\alpha\kappa)^2&\ddots\\&\ddots\ear,\notag\\&~~~~~~~~~~~~~~~~~ \tilde{h}^{\ell}_{-(E+3):E+3} = (z\!-\!\beta)\lba{c}0^{E+2}\\\alpha\kappa\\\alpha\!-\!z\\\alpha\kappa\\0^{E+2}\ear,\label{eq:Vh_l}\\
        \tilde{V}^{m^2}_{E+2\leftarrow E+1} &= \tilde{V}^{n^1}_{E+2\leftarrow E+1},~~ \tilde{h}^{m^2}_{-(E+2):E+2} = \tilde{h}^{n^1}_{-(E+2):E+2}.\label{eq:Vh_m2}
    \end{align}
\end{subequations}
\endgroup
Similarly, Theorem \ref{thm:OCP_IO}'s proof makes use of $\tilde{V}^q_{E^q\leftarrow E+1}$ and $\tilde{h}^q_{-E^q:E^q}$ for $q\in\{\gamma,\lambda,\psi,\omega\}$, which we derive by taking the inverse spatial transforms of \eqref{eq:IO_maps_of_lambda}.
\arraycolsep=1.3pt\def\arraystretch{0.8}
{\footnotesize
\begin{subequations}
    \begin{align}
        \tilde{V}^{\gamma}_{E+2\leftarrow E+1} &= \tilde{V}^{n^1}_{E+2\leftarrow E+1},~~ \tilde{h}^\gamma_{-(E+2):E+2} = 0^{2E+5},\\
        \tilde{V}^\lambda_{E+1\leftarrow E+1} &= I,~~ \tilde{h}^\lambda_{-(E+1):E+1} = 0^{2E+3},\\
        \tilde{V}^\psi_{E+3\leftarrow E+1} &= \tilde{V}^\ell_{E+3\leftarrow E+1},~~ \tilde{h}^\psi_{-(E+3):E+3} = \tilde{h}^\ell_{-(E+3):E+3},\\
        \tilde{V}^\omega_{E+2\leftarrow E+1} &= \tilde{V}^{n^1}_{E+2\leftarrow E+1},~~ \tilde{h}^\omega_{-(E+2):E+2} = 0^{2E+5}.
    \end{align}
\end{subequations}
}

\subsection{Extents of Closed-Loop Mappings}
In this subsection we examine the extent of the closed-loop maps for both SL and IO and show there is a nonzero bound on their extents for any internally stabilizing controller.\par
From \eqref{eq:SL_maps_of_r22}, by the number of applications of the $\undertilde{\sigma}$ operator,
\begin{subequations}\label{eq:SL_ext_r12}
    \begin{align}
        {\rm ext}(\underline{\tilde{R}}) = {\rm ext}(\underline{\tilde{N}}) &= 1 + {\rm ext}(\underline{\tilde{r}}^{12}),\\
        {\rm ext}(\underline{\tilde{M}}) = {\rm ext}(\underline{\tilde{L}}) &= 2 + {\rm ext}(\underline{\tilde{r}}^{12}).
    \end{align}
\end{subequations}
From \eqref{eq:SL_R22_expand}, for any causal $\underline{\tilde{f}}$ with extent $E<\infty$,
\be \label{eq:SL_ext_f}
    {\rm ext}(\underline{\tilde{r}}^{12}) = \max\{E,1\}.
\ee
From \eqref{eq:SL_ext_r12} and \eqref{eq:SL_ext_f},
\begin{subequations} \label{eq:SL_extent_bounds}
    \begin{align}
        {\rm ext}(\underline{\tilde{R}}) = {\rm ext}(\underline{\tilde{N}}) &= \max\{E+1,2\},\\
        {\rm ext}(\underline{\tilde{M}}) = {\rm ext}(\underline{\tilde{L}}) &= \max\{E+2,3\}.
    \end{align}
\end{subequations}
Similarly, for IO, from \eqref{eq:IO_maps_of_lambda} and \eqref{eq:IO_lambda_expand},
\begin{subequations} \label{eq:IO_extent_bounds}
    \begin{align}
        {\rm ext}(\underline{\tilde{\Gamma}}) = {\rm ext}(\underline{\tilde{\Omega}}) &= \max\{E+1,2\},\\
        {\rm ext}(\underline{\tilde{\Lambda}}) &= \max\{E,1\},\\
        {\rm ext}(\underline{\tilde{\Psi}}) &= \max\{E+2,3\}.
    \end{align}
\end{subequations}
Each closed-loop map for both SL and IO has a nonzero bound on its extent. $E$ being the extent of the decision variable $\tilde{f}$ rather than the extent of any specific closed-loop map, choosing $E=0$ does not actually reduce the extent of any of the closed-loop mappings compared to choosing $E=1$.

\end{document}